\documentclass[twocolumn,showpacs,preprintnumbers,amsmath,amssymb,superscriptaddress]{revtex4}

\usepackage{graphicx}
\usepackage{dcolumn}
\usepackage{bm}

\begin{document}


\title{Field-induced metal-insulator transition and switching phenomenon 
in correlated insulators}

\author{Naoyuki Sugimoto}
\affiliation{
Department of Applied Physics, University of Tokyo, Tokyo 7-3-1, Hongo, Tokyo 113-8656, Japan}
\email{sugimoto@appi.t.u-tokyo.ac.jp}
\author{Shigeki Onoda}%
\affiliation{%
Condensed Matter Theory Laboratory, RIKEN (The Institute of Physical 
and Chemical Research), Wako 351-0198, Japan
}%
\author{Naoto Nagaosa}
\affiliation{
Department of Applied Physics, University of Tokyo, Tokyo 7-3-1, Hongo, Tokyo 113-8656, Japan}
\affiliation{
Cross-Correlated Materials Research Group (CMRG), ASI,
RIKEN, 2-1 Hirosawa, Wako, Saitama 351-0198, Japan
}%

\date{\today}

\begin{abstract}
We study the nonequilibrium switching phenomenon associated with
the metal-insulator transition under electric field $E$
in correlated insulator by a gauge-covariant Keldysh formalism. 
Due to the feedback effect of the resistive current $I$, 
this occurs as a first-order transition with a hysteresis 
of $I$-$V$ characteristics having a lower threshold electric field
( $\sim 10^4 $Vcm$^{-1}$ ) much weaker than that for the Zener breakdown.
It is also found that the localized mid-gap states 
introduced by impurities and defects 
act as hot spots across which the resonant tunneling occurs selectively, 
which leads to the conductive filamentary paths and 
reduces the energy cost of the switching function. 
\end{abstract}

\maketitle

In correlated electronic systems, 
the Coulomb interaction and the electron-phonon coupling give rise 
to various long-range orderings of spin, charge, and orbital degrees 
of freedom of electrons, providing rich phase diagrams 
and intriguing phenomena such as colossal magneto-resistance~\cite{CMR}. 
The collective response can be significantly sensitive and amplified 
in comparison with that in semiconductors~\cite{SZE}, 
because many electrons cooperate in a short length scale of nanometres 
owing to the high electron density. 
These orderings often lead to an insulating behavior with 
an energy gap in the single-electron spectrum represented 
by the Mott gap~\cite{Mott}. 
In sharp contrast to the band insulators, 
the gap itself can be controlled by the external stimuli, 
as observed in experiments on the metal-insulator transition 
driven by the electric field~\cite{Asamitsu,Takubo} 
or the light irradiation~\cite{Miyano99,Miyano_prl97}. 
Furthermore, metal-insulator switching phenomena have been observed in 
other correlated electronic systems 
such as organic charge-transfer compounds~\cite{Kumai99}, 
La$_{2-x}$Sr$_x$NiO$_4$~\cite{Ni}, 
one-dimensional Mott insulators Sr$_2$CuO$_3$/SrCuO$_2$~\cite{SrCuO}. 
Recently, the application of the switching phenomenon to electronic devices has also been seriously considered~\cite{Sawa}. 
One important observation here is that 
the threshold electric fields observed in the correlated systems 
are typically  $10^4V/$cm~\cite{Ni,SrCuO} 
which is much less than $\sim 10^6 V/$cm 
expected from the simple Zener breakdown (see below). 
This suggests a positive feedback effect of the collective 
nature of the metal-insulator transition in the switching phenomena. 
Also the current is often non-uniform and confined 
in narrow paths or filaments~\cite{Takubo,Dagotto}. 

Theoretically, on the other hand, the description of 
the nonequilibrium states still remain a challenge even though 
there are several related works~\cite{Oka,OkamotoMillis,Okamoto}. 
Especially the non-perturbative treatment of the steady state 
under a strong electric field has been a difficulty. 
Recently, we have developed such formalism 
to deal with the far-from-equilibrium states~\cite{Onoda06_ptp}. 
This enables us to exactly incorporate the effects of the electric field into the Dyson equation for the nonequilibrium Green's function, which is written in a compact form by using the Moyal product 
in the gauge-covariant Wigner representation. 

In this Letter, we develop a theory of resistive switching 
phenomenon in the spin/charge ordered correlated insulators 
employing the gauge-covariant Keldysh formalism combined with 
the mean-field approximation to the electron-electron interaction. 
Then, the hysteric resistive switching due to the applied electric 
field has been obtained theoretically for the first time as far as we know. 
The theory also accounts the experimentally observed low threshold field 
and filament formation. 


We study the one-dimensional interacting electrons described by 
\begin{equation}
\hat H = \sum_{p \sigma} \varepsilon(p) 
c^\dagger_{p, \sigma} c_{p, \sigma}
+ g \sum_{p_1,p_2,q}c^\dagger_{p_1+q, \uparrow}
c^\dagger_{p_2-q, \downarrow} c_{p_2, \downarrow}
c_{p_1, \uparrow},
\end{equation}
where $g$ is assumed to be a constant and repulsive and the other notations are 
standard. This interaction naturally leads to the spin density wave ordering at the 
wavevector $2k_F$ ($k_F$: Fermi wavenumber), which is assumed to be half of 
the reciprocal lattice vector $G$, i.e., half-filling, and introduces the gap 
$2\Delta$. The sign of the gap is the opposite for the opposite spin, 
and we can just consider the two copies of the spinless electrons 
by the mean field Hamiltonian;
\begin{eqnarray}
\hat H\cong \sum _p{{\vec c}_p}^{\, \dagger }\left (
\begin{array}{cc}
vp & \Delta \\
\Delta & -vp
\end{array}
\right )
{\vec c}_p,\label{eq:Hamiltonian} 
\end{eqnarray}
where, $ {\vec c}_p= {}^t (c_{p,R},c_{p,L})$ is the two component 
operator corresponding to the right-going and left-going electrons 
near $\pm k_F$ with the dispersion $\pm vp$ ($v$: velocity). 
Here, $2\Delta :=g\overline {\langle c^\dagger _{p,R}c_{p,L}\rangle }$ 
is the self-consistently determined gap 
(see Eq.~(\ref{eq:self})). 
While this ordered state in equilibrium is well-known, 
we are interested in the nonequilibrium phase transition driven by the 
electric field. 
Note that the ordering is commensurate, 
and the phason degrees of freedom is quenched in sharp contrast to the 
sliding charge density wave problem \cite{CDW}. 
The interaction is treated in the mean-field approximation, which is justified in the weak to intermediate-coupling regime, 
though a more elaborate treatment is required in the strong-coupling regime. 

We separate the problem into two steps, i.e., (i) to 
describe the current flowing state under the electric field 
in the mean field Hamiltonian Eq.~(\ref{eq:Hamiltonian}), 
and (i\hspace{-.1em}i) to solve the self-consistent equation 
$2\Delta = g\overline {\langle c^\dagger_{p, R} c_{p, L}\rangle }$ for the gap. 
The first step is basically the Zener tunneling problem studied 
previously~\cite{Landau,Zener, Strueck,Gafen,Ao,Ziman}. 
As schematically shown in Fig.~1, 
the electrons tunnel through the energy gap. 
\begin{figure}
\begin{center}
\includegraphics[width=5cm]{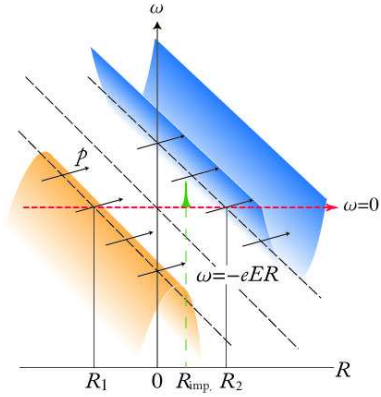}
\end{center}
\caption{The tilted band structure under an external electric field $E$. 
At each spatial position $R$, the momentum $p$ is 
defined as shown in the orthogonal direction. 
The conduction band bottom and the valence band top 
cross the energy $\omega=0$ at $R_2$ and $R_1$, respectively.  
The localized impurity state at $R=R _{\rm imp.}$ 
with green color gives rise to the resonant tunneling.}
\end{figure}
The band structure is spatially tilted by the potential energy 
gain $-eER$ where $R$ is the real space position and $-e$ is the electronic charge. 
One can consider the locally defined band structure as a function of the 
momentum (which is represented along the transverse axis) at 
each $R$, and the equi-energy line crosses the bottom (top) of 
the conduction (valence) band at $R=R_1$ ($R=R_2$). 
The wavefunctions of conduction and valence bands 
tunnel through the potential barrier between $R_1$ and $R_2$ 
from the both sides. 

There are three length scales with this problem; 
(i) the correlation length $\zeta = \hbar v/2\Delta $ 
associated with the energy gap $2\Delta $, which describes the 
characteristic extent of the wave packet relevant to the 
tunneling, (i\hspace{-.1em}i) the tunneling length 
$\xi =2\Delta /eE=R_2-R_1$ over 
which an electron can gain the energy $2\Delta $ by the electric 
field $E$, and (i\hspace{-.1em}i\hspace{-.1em}i) 
the mean free path $\ell $. 
The Zener tunneling occurs quite differently depending on the 
relative magnitudes of these length scales. 
We will focus below the case of $\ell \gg \zeta$; the mean 
free path is much longer than the correlation length, or the 
energy gap $2\Delta$ is much larger than the energy 
broadening $\hbar v/\ell$ due to the impurity scatterings. 
Then the Zener tunneling is controlled by the ratio of $\xi /\zeta $. 
When $\xi \gg \zeta $, the Zener tunneling probability 
can be calculated in the semi-classical approximation as 
$\sim \exp[ - \pi \xi /2\zeta ]$~\cite{Ziman}. 
As we increase the electric field so that $\xi <\zeta $, the wave  
packet extends from $R_1$ to $R_2$ and the metallic conduction 
occurs, i.e., the crossover between the Zener tunneling and 
Ohmic regions. 

Although this picture is valid qualitatively, 
it is crucial to consider steady state with the 
dissipative current flowing to describe the 
nonequilibrium phase transition. 
We perform the self-consistent calculations of Green's functions and self-energies in the Keldysh space in the gauge-covariant Wigner representation, which is now composed of the mechanical energy and momentum~\cite{Onoda06_ptp,Sugimoto}. 
It is necessary to introduce the Green's functions 
and the self-energies in the Keldysh space, 
$\underline G:=\left (
\begin{array}{cc}
\hat G^R&2\hat G^<\\
0&\hat G^A
\end{array}\right )$, 
and 
$\underline \Sigma :=\left (
\begin{array}{cc}
\hat \Sigma ^R&2\hat \Sigma ^<\\
0&\hat \Sigma ^A
\end{array}\right )$, respectively \cite{Onoda06_ptp}. 
Kinetic equations of the functions are given in the form 
of the Dyson equations: 
$(\hat {\cal L}-\underline \Sigma )\star \underline G=1$ and 
$\underline G\star (\hat {\cal L}-\underline \Sigma )=1$ 
with $\hat {\cal L}(\omega ,p):=\omega -\hat H(p)$. 
The symbol $\star $ denotes the Moyal product: 
$
(f\star g)(x)=\frac{1}{(\pi eE\hbar v)^2}\int dydzf(y)g(z)
e^{-\frac{2i}{eE\hbar v}((x^\mu -y^\mu )S_{\mu \nu }(x^\nu -z^\nu ))}
$, 
where $x,y,z$ denote the two-dimensional energy-momentum 
coordinates $(\omega ,vp)$, $f$ and $g$ are 
the smooth functions of the energy-momentum, and 
$S_{\mu \nu }:=
\left (\begin{array}{cc}0&-1\\1&0\end{array}\right )$. 

Now, we turn to the calculation of $\hat{G}^{R,A}$ and $\hat G^<$ 
for the Landau-Zener model with the $\delta$-functional random impurity potential. 
Let us start with the pure case of $\hat \Sigma ^{R,A}=\mp i\eta $ with an 
infinitesimal number $\eta $. 
By using a Fourier transform: 
${\cal F}[f](\omega ,\epsilon ):=\int \frac{dp}{2\pi \hbar }
f(\omega ,p)e^{-2ip\epsilon /\hbar eE}$, 
we obtain the Green's function in the pure case as 
\begin{eqnarray}
\hat G^{R,A}_{\rm pure}(\omega ,p)&=&\sum _{s=1,2}
\frac{2}{eE}\int d\epsilon d\Upsilon 
\frac{\hat \Phi _s(\Upsilon +\epsilon )
\hat \Phi ^\dagger _s(\Upsilon -\epsilon )}{\omega -\Upsilon \pm i\eta }
e^{\frac{2ip\epsilon }{eE\hbar }},\nonumber \\ \label{eq:greenpure}
\end{eqnarray}
where 
$\hat \Phi _s:= {}^t~\left (\Phi ^+_s, \Phi ^-_s \right )$ 
are the solutions of the following Weber equations~\cite{Zener}:
$
\left (\partial _z^2+\frac{1}{2}
\pm i\frac{\Delta ^2}{2eE\hbar v}-
\frac{z^2}{4}\right )\Phi ^\pm _s\left (\sqrt\frac{eE\hbar v}{2}
e^{\mp \frac{i\pi}{4}}z\right )=0 
$
with a normalization condition: 
$
\sum _{s=1,2}\int d\lambda \hat \Phi (\lambda -\omega )
\hat \Phi ^\dagger (\lambda -\omega ^\prime )
=\delta ((\omega -\omega ^\prime )/eE)$. 
We employed the self-consistent Born approximation; 
$\hat \Sigma ^{R,A}(\omega )=n_iu^2\hat g^{R,A}(\omega ,\epsilon =0)$, 
with the density of impurities $n_i$ and the strength of the potential $u$ 
leading to the lifetime $\tau :=(v\hbar ^2/n_iu^2)$ and 
the mean-free path $\ell :=\tau v$. 
Then, we calculate the retarded and advanced Green's functions through
$\hat G^{R,A}=
\hat G^{R,A}_{\rm pure}+\hat G^{R,A}_{\rm pure}\star \hat \Sigma ^{R,A}\star \hat G^{R,A}$. 
Finally, the lesser Green's function is obtained as 
${\hat G}^<\cong f_F\star {\hat G}^A-{\hat G}^R\star f_F+
{\hat G}^R\star [f_F\star 
\begin{picture}(0,0)
\put(-6,-3){,}
\end{picture}
\hat {\cal L}]\star {\hat G}^A$
with the Fermi distribution function $f_F(\omega )$. 
Here, we neglect a vertex correction since it gives only a minor correction~\cite{Rickayzen}. 

The self-consistent mean-field gap equation of 
$\Delta$ for a given interaction strength $g$ is given by 
\begin{equation}
{ 1 \over g} \Delta =\overline {\langle c^\dagger _{p,R}c_{p,L}\rangle }/2
\equiv \int \frac{d\omega }{4i\pi }\frac{dp}{2\pi \hbar }{\rm tr}\left [
\hat \sigma ^x\hat G^<(\omega ,p)\right ],\label{eq:self}
\end{equation}
the right-hand side of which is a function of the 
electric field $E$ and the gap $\Delta$ itself. 
Throughout this paper, we take $g=5\hbar v$. 
The solution is obtained by the crossings of the straight line $\Delta/g $ 
and the curve for $\overline {\langle c_{p,L}^\dagger c_{p,R}\rangle }/2$ in Fig.~2(a). 
\begin{figure}
\begin{center}
\includegraphics[width=8cm]{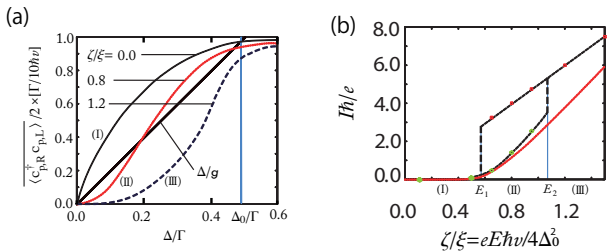}
\end{center}
\caption{
(a): The right-hand side of the gap equation Eq.~(\ref{eq:self}), 
$\overline {\langle c^\dagger _{p,R}c_{p,L}\rangle }/2$ in the unit of 
$(\Gamma /10\hbar v)$ as a function of the gap 
$\Delta$ for $\zeta /\xi =eE\hbar v/4\Delta _0^2=0.0, 0.8, 1.2$,  
with $\Gamma $ being the energy cutoff, i.e., 
the half bandwidth of the equilibrium states. 
Here, $\Delta _0$ is the gap in the equilibrium for $g=5\hbar v$. 
The solid straight line represents $\Delta /g$ in the same unit 
and the crossing of these two gives the solution(s) 
to the mean field equation (\ref{eq:self}). 
(b): The obtained current $I$ as a function of the electric field $E$. 
The dashed line is a guide to the eyes. 
The $I$-$E$ characteristics clearly shows the hysteresis and 
switching behavior of the current. 
The red dotted curve represents the current obtained for the fixed gap $\Delta _0$. 
}
\end{figure}
There exist three regions of the strength of the electric field as 
(I) $E<E_1 $, (I\hspace{-.07em}I) $E_1<E<E_2$, and (I\hspace{-.07em}I\hspace{-.07em}I) $E>E_2$, 
where the number of solutions to Eq.~(\ref{eq:self}) is 
two, three, and one, respectively. 
Note that the stability of each solution is determined by the condition 
$\partial \left (\overline {\langle c^\dagger _{p,R}c_{p,L} \rangle }/2 \right )
/\partial \Delta <1/g$. 
Thus, in region (I), the finite-$\Delta$ solution is the only stable one. 
In Fig. 2(a), we show the case of the equilibrium ($E=0$) with the gap $\Delta _0$. 
In the region (I\hspace{-.07em}I), 
there are two stable solutions, 
i.e., $\Delta=0$ and $\Delta \ne 0$ 
as shown for the case of $\zeta /\xi =eE\hbar v/4\Delta ^2_0=0.8$ in Fig.~2(a), 
except the intermediate unstable one. 
This is the typical situation of the first-order phase transition. 
In the region (I\hspace{-.07em}I\hspace{-.07em}I), 
the stability of the $\Delta \ne 0$ solution is lost, and the 
metallic state ($\Delta=0$) becomes the only stable solution, 
as shown for the case of $\zeta /\xi =1.2$ in Fig.~2(a). 
Therefore, we conclude that the spin/charge ordered 
system shows the first-order-like switching phenomenon. 

We argue that these two threshold electric fields 
are essentially given by 
$E_1=\Delta _0/ev\tau $ and $E_2=\Delta ^2_0/e\hbar v$. 
It is easy to understand that $E_2$ is the Zener breakdown field 
since at $E>E_2$, the gap does not prevent the metallic current 
flow and hence the insulating state is unstable. 
To understand why the lower threshold field $E_1$ appears, 
it is useful to consider the instability of 
the metallic current-carrying state. 
The steady state with the current is characterized by 
the shift of the electron distribution function by the amount 
$\delta k = e E \tau$ with $\tau =\ell/v$ being the mean-free time. 
With this shift, the energy difference between the right and 
left-moving electrons at the shifted Fermi level is 
$\delta \varepsilon = 2 v \delta k$. 
When this energy is larger than the gap $2 \Delta_0$ 
in the equilibrium state, the instability toward the SDW/CDW disappears. 
This consideration leads to the estimation 
$E_1=\Delta _0/ev \tau $, which is smaller than the 
Zener breakdown field $E_2=\Delta ^2_0/e\hbar v$ 
by the factor $\hbar/ (\tau \Delta _0) \ll 1$. 

Now we study the physical properties associated with the 
switching phenomenon. 
Of the most important is the $I$-$E$ characteristics. 
The current $I$ flowing through the sample is obtained 
from the relation 
\begin{eqnarray}
I=\frac{e^2v^2E}{2}\int \frac{d\omega }{2\pi }
\frac{dp}{2\pi }{\rm tr}\left[
\hat \sigma ^z\hat G^R\star \left (
-\hat \sigma ^z\frac{\partial f_F}{\partial \omega }
\right )\star \hat G^A
\right ]. \label{eq:I}
\end{eqnarray}
Figure 2(b) shows the $I$-$E$ characteristics corresponding to the 
first-order phase transition of the order parameter obtained in Fig.~2(a). 
There occurs the jump of the current, the upper branch of which corresponds 
to the metallic conduction while the lower branch to the Zener tunneling 
in the insulating state. The two threshold electric fields $E_1$ and $E_2$ 
can be separated by a factor as discussed above, and the 
change of the current is by the factor of $\sim {\rm exp}(4\Delta^2/e\hbar v E)$. 

We also propose the measurement of the local density of states (LDOS) 
in terms of the scanning tunneling spectroscopy (STS) to study the 
nonequilibrium state. Based on the formula for the tunneling current 
given by Meir and Wingreen~\cite{Meir}, we have 
calculated the STS LDOS as shown in Fig.~3. 
There appears a peak at the middle of the 
gap whose height is proportional to the 
tunneling current. 
Namely, the tunneling occurs through the in-gap density of states 
induced by the electric field. 
In the metallic state after the 
switching, of course the gap completely closes. 
\begin{figure}
\begin{center}
\includegraphics[width=6cm]{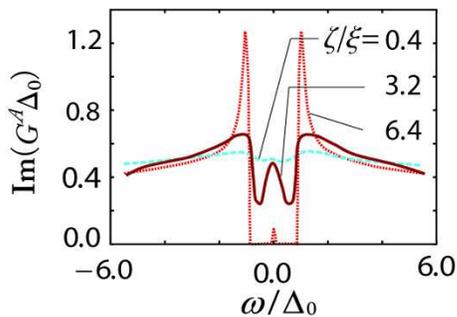}
\end{center}
\caption{
The local density of states (LDOS) ${\rm Im}\left (\hat G^A\Delta _0\right )$ 
as a function of the normalized energy $\omega /\Delta _0$ 
($\Delta _0$: the energy gap in the equilibrium) 
for $\zeta /\xi $=6.4 (dashed line), 
3.2 (solid line) and 0.4 (dotted line), respectively. 
($\zeta=\hbar v/2\Delta _0$: correlation length, $\xi= 2 \Delta _0/eE$) 
}
\end{figure}

Now the semi-quantitative estimation for the realistic situation is in order. 
Typically, $2\Delta$ is of the order of $1eV$, 
while $\hbar/\tau$ is $\sim$ 10m$eV$, 
which leads to the factor of $\sim 100$ reduction 
of the switching threshold from the Zener breakdown field $\sim 10^6 V/$cm. 
Therefore, the observed values of the order of $10^4V/$cm\cite{Ni,SrCuO} 
is in the reasonable range as expected from the present consideration. 
A threshold current density $j_1$ is estimated by 
$j_1\sim (\Delta na)(e/\hbar )$, 
where $n$ denotes a density of the electron. 
By using the typical values, we estimate the current density 
as $j_1\sim 10^8{\rm A/cm}^2$. 
This is a very large current density, 
and can be usually realized only in the pulse current 
experiment since the huge heat generation makes the sample burned out. 

However, the breakdown and switching  occur often in the 
filamentary paths of metallic regions \cite{Takubo,Dagotto}. 
As mentioned above, the emerging in-gap state is 
associated with the tunneling. In real materials, 
there are often in-gap states due to the impurities, vacancies, 
{\it etc} even without the electric field. 
Suppose there is an impurity level in the gap as shown 
by the green peak at $R=R_{\rm imp.}$ in Fig.~1. 
Let $t_1$ be the tunneling amplitude from the valence 
band at $R=R_1$ to the impurity level, 
while $t_2$ be that from there to the conduction band at $R=R_2$. 
Note that $t_1$ and $t_2$ are exponentially small, 
i.e., $t_1 \sim {\rm exp}( - \pi (R_{\rm imp.}-R_1)/2 \zeta)$, 
$t_2 \sim {\rm exp}( - \pi (R_2-R_{\rm imp.})/2 \zeta)\ll 1$, 
with the product $t_1 t_2$ being the Zener tunneling amplitude
without the resonant level. 
Considering the two barrier problem with the small tunneling amplitudes of $t_1$ and $t_2$, the tunneling probability through the two barriers has the peak height given by $T_{\rm max} \cong [2 t_1 t_2/(t_1^2 + t_2^2)]^2$ within the narrow energy width $\delta \varepsilon \cong W(t_1^2+ t_2^2)$, which can be translated into the width in the real-space 
$\delta R \cong \delta \varepsilon /(eE)$ in Fig.1.
When the extent of the electron wave packet is larger 
than $\delta R$ (which is the case in the limit of small tunneling amplitude), 
the averaged tunneling probability is of the order of 
$t_1^2 t_2^2/(t_1^2+t_2^2)$. 
This is much larger than that without the impurity level, i.e.,
$(t_1 t_2)^2$ corresponding to the tunneling amplitude  $t_1 t_2$
for the Zener tunneling. 
This means that there appear ``hot spots'' at 
$R_1$ and $R_2$ which are spatially separated by $\zeta$. 
Therefore the impurities act as the nucleation centers of the 
nonequilibrium first-order metal-insulator phase transition 
discussed above. 
With the random configuration of the impurities, 
the current can find the path along where this nucleation centers 
populate densely compared with the other spatial region. 
This leads to the filamentary paths of the metallic regions as 
observed experimentally. 

Since the width of the filamentary path is given by the correlation length $\zeta$, which characterizes also the spatial variation of the SDW/CDW order parameter, 
a threshold current $I_1$ is estimated by $I_1\sim \pi \zeta ^2\cdot j_1=
(na v^2/\Delta )(\pi e\hbar )\sim 1\mu $A. 
From this expression, we found that the switching occurs 
with rather tiny current in correlated insulators. 
Moreover, from a point of view of the 
Joule heating, the correlated insulator is 
more advantageous over semiconductors. 
The Joule heat corresponding to the breakdown is written as 
$P=j_1\cdot E_1=(\Delta a)^2(1/\hbar v\tau )$. 
For the correlated insulator, this 
is estimated as $P_{\rm cor.}\sim 10^{12}{\rm VA/cm}^3$. 
This value is rather similar to that in the typical semiconductors. 
However, as mentioned above, the current is confined within 
filaments of nano-scale in the correlated systems, 
while it is rather uniformly distributed in semiconductors. 
Therefore, the total heat generation $ P_{\rm cor.}^\prime$ 
is expected to be 
much smaller in correlated insulator as 
$P_{\rm cor.}^\prime =\varrho \pi \zeta ^2 j\sim \varrho 
\times 10^{-2}{\rm VA/cm} ^3$, 
where $\varrho $ denotes a density of the filaments. 
Once some filaments appear, the voltage drop 
across the sample disappears and no additional 
filaments are needed. 

To summarize, we have studied the switching phenomenon 
of the correlated insulator under an applied electric field. 
Due to the feedback effect of the current on the spin/charge ordering, 
the switching occurs with much weaker field/smaller current/
smaller heat generation as compared with those expected 
from the simple Zener breakdown picture. 
This finding will be useful for the future application of this phenomenon 
to memory and switching devices. 

S.~O. thanks S. Okamoto for discussion. 
The work was partly supported by Grant-in-Aids 
(No. 15104006, No. 16076205, No. 17105002, No. 19048015) 
and NAREGI Nanosicence Project from the Ministry of Education, Culture, Sport, 
Science and Technology. 
S.~O. was supported by Grant-in-Aids (No. 19840053) from 
Japan Society of the Promotion of Science.

\end{document}